\documentclass[11pt]{article}

\usepackage{amsfonts}
\usepackage{amsmath}
\usepackage{graphicx}

\newcommand{\baryprec}{\mathcal{P}}
\newcommand{\causet}{C}
\newcommand{\coevcomp}{\mathcal{U}}
\newcommand{\coevent}{\kappa}
\newcommand{\future}{J_+}
\newcommand{\hplane}{H}
\newcommand{\incmatr}{\mathbf{I}}
\newcommand{\linhull}{\mathcal{L}}
\newcommand{\rfield}{\mathbb{R}}
\newcommand{\timelet}{T}
\newcommand{\timelets}{\mathcal{T}}
\newcommand{\totalsimplex}{\mathcal{S}}

\newtheorem{statement}{Statement}

\DeclareMathOperator{\diag}{diag}
\DeclareMathOperator{\supp}{supp}

\title{Timelets on causal sets}

\author{Rom\`an Zapatrin}

\begin{document}

\maketitle

\begin{abstract}
Dual structures on causal sets called timelets are introduced, being discrete analogs of global time coordinates. Algebraic and geometrical features of the set of timelets on a causal set are studied. A characterization of timelets in terms of incidence matrix of causal set is given. The connection between timelets and preclusive coevents is established, it is shown that any timelet has a unique decomposition over preclusive coevents. The equivalence classes of timelets with respect to reascaling are shown to form a simplicial complex. 
\end{abstract}

\section*{A foreword}

Causal set approach is intended to represent spacetime structure as a discrete one. It has promising application in both quantum gravity and foundations of quantum mechanics. On the other hand, causal sets may be treated as just Regge-like approximations of spacetime, which is thought of as being `indeed continuous'. This paper has a technical character, it has nothing to do with interpretation, I just introduce a dual structure on causal sets. Roughly speaking, the elements of a causal set (causet in the sequel) are events, while timelets I introduce are dual structure on the set of events, respecting the causal order. Timelets generalize the notion of coevents \cite{Sorkin2007}, sequential growth parameter \cite{RideoutSorkin99}, they look like analogs of global time. 

The paper is organized as follows. First, the definition of timelets is provided along with their matrix representation and the monotonicity condition in terms of the incidence matrix is formulated. The classes of monotonically equivalent timelets are described and the representation of generic timelets as maximal extensions of the partial order on the causet is provided. In section \ref{spreclcoev} the most degenerate, two-valued timelets are described, they are associated with preclusive coevents and the decomposition of timelets over preclusive coevents is provided. Then the equivalence classes of timelets are shown to form simplicial complexes, two examples are elaborated in detail. 

\section{The cone of timelets}\label{sdef}

\paragraph{Definition of timelet.} Given a causet $\causet$ and any pair $a,b\in \causet$, a \emph{timelet} on $\causet$ a function $\timelet$ satisfying
\begin{equation}\label{defmon}
a\le b \;\Longrightarrow \; \timelet(a)\le \timelet(b)
\end{equation}

Let $\causet$ be a (finite) causal set. Denote by $\linhull$ its linear span -- a vector space whose basis is labelled by the elements of $\causet$. Any element $\timelet\in \linhull$ has its definite coordinates, that is, it assigns certain number to each event. Although, such assignment does not capture the causal structure of $\causet$. Only the elements $\linhull$ respecting the causal order are called timelets. 

Denote the set of all timelets on $\causet$ by $\timelets\left(\causet\right)$ or just $\timelets$ if no ambiguity occurs. 
The set $\timelets$ is a cone, that is, it is closed under multiplying by positive numbers and by taking convex combinations: for any $C,C'\in \timelets$
\[
\begin{array}{l@{\qquad}l}
\kappa C \in \timelets & \mbox{for any $\kappa>0$}
\\
\lambda C + (1-\lambda C') \in \timelets & \mbox{for any $\lambda\in [0,1]$}
\end{array}
\]

The set of timelets form a convex subset $\timelets$ of the $\linhull$, let us describe it. This description follows directly from the definition \eqref{defmon}, namely
\begin{equation}\label{defhp1}
\timelets=
\left\{
\tau\in\totalsimplex
\mid
\forall x,y \in \causet 
\qquad
x<y
\;\Rightarrow\;
\timelet(x)\le \timelet(y)
\right\}
\end{equation}
It follows from the above that the set $\timelets$ of all timelets is a polyhedron delimited by the hyperplanes \eqref{defhp1}. Furthermore, due to the transitivity of partial order there are redundant hyperplanes. The least sufficient set of conditions looks similar, but refers to covering relation $\prec$ instead of partial order $\le$. The covering relation is defined as follows:
\begin{equation}\label{defprec}
\left\lbrace
\begin{array}{l}
x<y
\\
\quad
\\
\forall z \mbox{ if } x\le z \le y \mbox{ then } z=x \mbox{ or } z=y
\end{array}
\right.
\end{equation}
So, the necessary set of supporting hyperplanes is exhausted by those of the form\footnote{Another argument for that is that a real-valued function on a partially ordered set is monotone iff it is monotone on any 3-chain \cite{Burkill1984}}
\begin{equation}\label{defhp2}
\forall x,y \in \causet 
\qquad
x\prec y
\;\Rightarrow\;
\timelet(x)\le \timelet(y)
\end{equation}
All supporting hyperplanes are similar, they are labelled by pairs $\{(x,y) \mid x\prec y\}$. Each such hyperplane $\hplane_{(x,y)}$ is defined by the equation
\[
\hplane_{(x,y)}=\{\tau \in \linhull \mid \tau(x)=\tau(y)\}
\]

\section{Matrix representation of timelets}\label{smatrep}

Let $N$ be the cardinality of the causet $\causet$. Consider the set of all \emph{diagonal} $N\times N$ matrices. Any function $\timelet: \causet \to \rfield$ can be represented by such a matrix. 
\[
\timelet_{xy}
\;=\;
\begin{cases}
\timelet(x), & \mbox{if $x=y$}
\\
0 & \mbox{otherwise}
\end{cases}
\]
No ambiguity occurs if the same symbol $\timelet$ is used for both timelet and the matrix representing it. Let us derive an algebraic condition for the matrix representing $\timelet$ to be a timelet. The possibility to extend functions $\rfield\to\rfield$ to functions $\timelets\to\timelets$ is an advantage of matrix representation. 

\paragraph{Incidence matrix.} For a partially ordered set, the matrix whose entries labelled by pairs of elements of $\causet$ having the form
\begin{equation}\label{eincmatr}
\incmatr(x,y)
\;=\;
\begin{cases}
1, & \mbox{if $x\le y$}
\\
0 & \mbox{otherwise}
\end{cases}
\end{equation}
Any partially ordered set can be fully ordered, therefore the matrix $\incmatr$ can be made upper-diagonal using appropriate enumeration of the causet $\causet$. For instance, for the causet $\causet$ 
\unitlength1mm
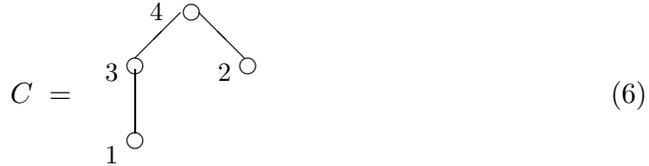
\begin{equation}\label{eex1}
\causet
\;=\;
\raisebox{-0.7cm}{
\begin{picture}(30,22)(0,3)
\put(5,5){\circle{2}}
\put(1,2){\mbox{\small{1}}}
\put(1,13){\mbox{\small{3}}}
\put(16,13){\mbox{\small{2}}}
\put(7,21){\mbox{\small{4}}}
\put(5,6){\line(0,1){8.5}}
\put(5,15){\circle{2}}
\put(20,15){\circle{2}}
\put(5,16){\line(1,1){6}}
\put(19.5,16){\line(-1,1){6}}
\put(12.5,22){\circle{2}}
\end{picture}
}
\end{equation}
\bigskip the appropriate incidence matrix has the form 
\begin{equation}\label{eexincmatr}
\incmatr
\;=\;
\left(
\begin{array}{cccc}
1&0&1&1
\\
0&1&0&1
\\
0&0&1&1
\\
0&0&0&1
\end{array}
\right)
\end{equation}

\begin{statement}[Monotonicity condition.] Let $\causet$ be a causet, $T$ be a diagonal matrix representing a function $\causet \to \rfield$. Then $T$ is monotone with respect to the partial order on $\causet$ if and only if
\begin{equation}\label{emoncond}
[\incmatr,\timelet] \ge 0
\end{equation}
\end{statement}
Calculate the entries of the matrix $M=[\incmatr,\timelet]$:
\[
M_{xy}
\;=\;
\begin{cases}
\timelet(y)-\timelet(x), & \mbox{if $x\le y$}
\\
0 & \mbox{otherwise}
\end{cases}
\]
which is a reformulation of the monotonicity condition.

\section{Timelets and preclusive coevents}\label{spreclcoev}

Timelets were announced to be analogs of global time on causal sets. In this section I introduce he analog of local coordinate transformations and study the connection of timelets with coevents -- dual structures on causets \cite{Sorkin2007}.

\paragraph{Monotonicity gauge.} Define the equivalence relation on the set of all timelets as follows. Let $\timelet_1$, $timelet_2$ be two timelets represented by appropriate matrices $\timelet_1$, $\timelet_2$. Then
\begin{equation}\label{emonequiv}
\timelet_1\sim\timelet_2
\mbox{ iff } \timelet_1=F(\timelet_2)
\end{equation}
where $F(\cdot)$ is a strictly monotonic function $\rfield\to\rfield$:
\[
x<y \; \Leftrightarrow \; F(x)<F(y)
\]
Intuitively, the monotonicity gauge is a local rescaling of clock at every point in accordance with the global causal structure. 

Let $\timelet$ be a generic timelet, that is, taking different values on the elements of the causet $\causet$. Using this equivalence relation \eqref{emonequiv}, the equivalence classes can be characterized by extension the partial order on $\causet$ to a full order. The conclusion is the following:
\begin{statement}
The equivalence classes of timelets with respect to the relation \eqref{emonequiv} are in 1--1 correspondence with enumerations $0,1,\ldots,N$ of the causet $\causet$ which respects its partial order. 
\end{statement}

First I considered generic timelets taking all different values on $\causet$. The other extreme case is when a timelet $\timelet$ takes only two values 0 and 1. Given a causet $\causet$ of cardinality $N$, coevents are a dual structure on it, bringing each element of $\causet$ into $\{0,1\}$. Denote the set of all coevents on $\causet$ by $\coevcomp_0$. In fact, since $\causet$ is finite, $\coevcomp_0$ is simply isomorphic to the power set of $\causet$
\[
\coevcomp_0
\simeq
2^\causet
\]
Using matrix representation, each coevent can be written down as a diagonal matrix $\diag(0011\ldots 1011)$. Let us take into account the partial order $\le$ defined on the causet $\causet$. Given a subset $X\subseteq \causet$, denote by $\future{X}$ its future
\[
\future(X)=
\left\{
y\in \causet 
\mid
\forall x\in X \;\; x\le y
\right\}
\]
In the meantime, any subset of $\causet$ is a coevent, so the mapping $\future(\cdot)$ acts on coevents
\[
\future: \coevcomp_0 \rightarrow \coevcomp_0
\]

\begin{statement}
The operation $\future$ is an idempotent operation on the set of coevents: $\forall X\in \coevcomp_0$
\[
\future \left(\future(X)\right)=\future(X)
\]
\end{statement}

\paragraph{Preclusive coevents.} A coevent $\coevent$ is \emph{preclusive} if its value on any precluded event is 0 \cite{Sorkin2007}. Zero value of a coevent $\coevent$ on a particular event $x$ means that $x$ is precluded. Since the ancestors of $x$ are also precluded, the requirement of preclusivity for the coevent $\coevent$ means
\begin{equation}\label{monprec}
\timelet(x)=0 \;\Rightarrow \quad \forall y\le x \quad \timelet(y)=0
\end{equation}
Denote the set of preclusive coevents by $\coevcomp$ on a caset $\causet$ by $\coevcomp(\causet)$, or simply $\coevcomp$ if no ambiguity occurs.
It follows immediately from the above that preclusive coevents viewed as functions on $\causet$ are monotone and therefore are timelets. 
\begin{statement}
Preclusive coevents, and only they, are invariant with respect to the operation $\future$
\begin{equation}\label{epreclcoev}
\coevcomp=\left\{X\in \coevcomp_0 \mid \future(X)=X \right\}
\end{equation}
or using the condition \eqref{emoncond} and the monotonicity gauge \eqref{emonequiv}
\begin{equation}\label{epreclinc}
X\in \coevcomp \; \Leftrightarrow \; [\incmatr,X] \sim X
\end{equation}
\end{statement}

\section{Simplicial representation of timelets}\label{sgeom}

In this section I introduce an analog of Lorenz transformation: time shift and boost. This make it possible to represent equivalence classes of generic timelets as simplices, so the set of all timelets acquire the structure of simplicial complex. 

\paragraph{Affine reduction of timelets.} Let us introduce an equivalence relation $\simeq$ on timelets being stronger than themonotonicity gauge $\sim$ defined in \eqref{emonequiv}. Rescaling means affine transformation: for any real $A$ and positive $B$.
\[
\timelet \equiv \timelet' 
\;\Leftrightarrow\;
\timelet'=A+B\timelet
\]
Adding appropriate constant and rescaling timelets by appropriate positive multiple we can always normalize them to fit into a simplex $\totalsimplex$ spanned on unit vectors in $\linhull$ and satisfy the following property
\begin{equation}\label{eaffeq}
\left\lbrace
\begin{array}{l}
\sum_{x\in \causet}\limits \timelet(x)=1 
\vspace{.7ex}
\\ 
\forall x\in \causet \quad x\ge 0
\\
\exists x\in \causet \quad x= 0
\end{array}
\right.
\end{equation}
It follows from \eqref{eaffeq} that any such timelet is inside the total simplex $\totalsimplex$. Furthermore, considering its barycentric coordinates, we see, as it follows from the third condition \eqref{eaffeq} that it is always on a face of $\totalsimplex$ rather than in its interior (because its minimal value is 0). The support of timelets is concentrated on simplices, which are faces opposite to vertices asoociated with minimal elements of the causet $\causet$. 

\paragraph{Decomposition of timelets.} Consider a generic timelet. Normalize it in such way that it has all different values on the elements of the causet $\causet$ and the minimal value is 0, the other values are positive integers. It follows immediately that the minimal zero value is the least, and it is achieved on a particular element $x_0\in\causet$. Consider the coevent $\coevent_0$ such that its support is the support of the timelet $\timelet$
\[
\coevent_1(x)
\;=\;
\begin{cases}
0, & \mbox{if $x=x_0$}
\\
1 & \mbox{otherwise}
\end{cases}
\]
The value 1 is achieved on a unique element $x_1\in\causet$ (since all the values of $\timelet$ are different). Define
\[
\timelet_1
\;=\;
\timelet_0 - \coevent_1
\]
$\timelet_1$ is a timelet whose all non-zero values are positive and different; however it is not normalized by the sum of its elements. Let $a_2 = \min_{x\in\supp(\timelet_1)} \timelet_1(x)$ achieved on an element $x_2\in\causet$. Introduce
\[
\coevent_2(x)
\;=\;
\begin{cases}
0, & \mbox{if $x\not\in\supp(\timelet_1)$}
\\
1 & \mbox{otherwise}
\end{cases}
\]
and define
\[
\timelet_2
\;=\;
\timelet_1 - \coevent_2
\]
and so on. Finally we get
\[
\timelet_{N-1}=\coevent_{N-1}
\]
where $N$ is the cardinality of the causet $\causet$ and the initial timelet $\timelet$ decomposes into a sum
\begin{equation}\label{edect}
\timelet
\;=\;
\sum_{k=1}^{N-1}
\coevent_k
\end{equation}
Each step of this decomposition is unique and unambiguous, hence the whole decomposition \eqref{edect} is unique. 

\medskip

That is, any generic timelet can be uniquely decomposed in this way into a convex combination of exactly $N-1$ preclusive timelets.

\paragraph{The total simplex.} Given the linear span $\linhull$ of the causet $\causet$, consider the simplex $\totalsimplex$ spanned on unit vectors associated with elements of $causet$, call it \emph{total simplex}
\[
\totalsimplex
\;=\;
\{ x\in\rfield^{N}
\mid
x_i\ge 0, \sum x_i=1
\}
\] 
It follows from the above that each equivalence class \eqref{eaffeq} is associated with a unique point of $\totalsimplex$, but not the vice versa. In the rest of the paper we explore the geometry of the set of so-normalized timelets. It follows from the third condition in \eqref{eaffeq} that the points of $\totalsimplex$ associated with normalized timelets reside on the faces of $\totalsimplex$ of dimension up to $N-1$, where $N$ is the cardinality of the causet $\causet$. Viewing coevents as real-valued (not necessarily monotone) functions on $\causet$ provides their natural representation in terms points of the simplex $\totalsimplex$. 

\paragraph{Barycentric representation of coevents.} Using the affine transformation \eqref{eaffeq} any coevent can be normalized as a timelet and given the form like $00\frac{1}{N}\frac{1}{N}\ldots \frac{1}{N}0\frac{1}{N}\frac{1}{N}$. 
\begin{statement}
There is 1--1 correspondence between nonzero coevents and faces (and barycenters as well) of the simplex $\totalsimplex$.
\end{statement}
In fact, any coevent can be transformed into a point of the form $00\frac{1}{N}\frac{1}{N}\ldots \frac{1}{N}0\frac{1}{N}\frac{1}{N}$, which is a barycenter of the appropriate face of the simplex $\totalsimplex$, while the faces of a simplex are comletely determined by their barycenters. Using this, from now on we will write for brevity 
\begin{equation}\label{esieq}
0011\ldots 1011 \quad \simeq \quad 00\frac{1}{N}\frac{1}{N}\ldots \frac{1}{N}0\frac{1}{N}\frac{1}{N}
\end{equation}
For a given normalized coevents $\coevent$, call the set of elements of the causet $\causet$ on which $\coevent$ takes nonzero values the \emph{support} of $\coevent$:
\begin{equation*}\label{esieq1}
\supp\left( 0011\ldots 1011 \vphantom{\int} \right)\; = \; \supp \left( 00\frac{1}{N}\frac{1}{N}\ldots \frac{1}{N}0\frac{1}{N}\frac{1}{N}\right)
\end{equation*}
Preclusive coevents are also have barycentric representation, denote the set of points of $\totalsimplex$ associated with preclusive by $\baryprec$

\medskip

\paragraph{Simplicial representation of timelets.} So, every coevent is (associated with) a barycenter of some face of the total simplex $\totalsimplex$. Any convex combination of timelets (and, in particular, preclusive coevents) is, in turn, a timelet. As it follows from the above, any timelet with nonnegative different values starting with zero can be uniquely decomposed over $N-1$ points in the total simplex $\totalsimplex$. So, all equivalence (with respect to the monotonic gauge) classes of timeles form a simplicial complex in the space $\linhull$ spanned on the causal set $\causet$.

\section{Examples} 

In this section I provide a detailed analysis of the structure of complexes of timelets on small causal sets. 

\paragraph{Example 1.} Return to the causet $\causet$ considered above (the numbers next to the vertices are just labels) and write down the decomposiition of generic timelets on $\causet$ in detail.
\begin{equation}\label{eex1e}
\causet
\;=\;
\raisebox{-0.7cm}{
\begin{picture}(30,22)(0,3)
\put(5,5){\circle{2}}
\put(1,2){\mbox{\small{1}}}
\put(1,13){\mbox{\small{3}}}
\put(16,13){\mbox{\small{2}}}
\put(7,21){\mbox{\small{4}}}
\put(5,6){\line(0,1){8.5}}
\put(5,15){\circle{2}}
\put(20,15){\circle{2}}
\put(5,16){\line(1,1){6}}
\put(19.5,16){\line(-1,1){6}}
\put(12.5,22){\circle{2}}
\end{picture}
}
\end{equation}
For this causet, the following nondegenerate covents are preclusive (filled circle corresponds to value 1 on the appropriate element of $\causet$), here is the complete list of preclusive coevents -- the elements of $\baryprec(\causet)$:
\unitlength1mm
\[
0111
=
\raisebox{-0.7cm}{
\begin{picture}(25,22)(0,3)
\put(5,5){\circle{2}}
\put(5,6){\line(0,1){8.5}}
\put(5,15){\circle*{2}}
\put(20,15){\circle*{2}}
\put(5,16){\line(1,1){6}}
\put(19.5,16){\line(-1,1){6}}
\put(12.5,22){\circle*{2}}
\end{picture}
}
;\;
1011
=
\raisebox{-0.7cm}{
\begin{picture}(25,22)(0,3)
\put(5,5){\circle*{2}}
\put(5,6){\line(0,1){8.5}}
\put(5,15){\circle*{2}}
\put(20,15){\circle{2}}
\put(5,16){\line(1,1){6}}
\put(19.5,16){\line(-1,1){6}}
\put(12.5,22){\circle*{2}}
\end{picture}
}
\]
\[
0101
=
\raisebox{-0.7cm}{
\begin{picture}(25,22)(0,3)
\put(5,5){\circle{2}}
\put(5,6){\line(0,1){8.5}}
\put(5,15){\circle{2}}
\put(20,15){\circle*{2}}
\put(5,16){\line(1,1){6}}
\put(19.5,16){\line(-1,1){6}}
\put(12.5,22){\circle*{2}}
\end{picture}
}
;\;
0101
=
\raisebox{-0.7cm}{
\begin{picture}(25,22)(0,3)
\put(5,5){\circle{2}}
\put(5,6){\line(0,1){8.5}}
\put(5,15){\circle*{2}}
\put(20,15){\circle{2}}
\put(5,16){\line(1,1){6}}
\put(19.5,16){\line(-1,1){6}}
\put(12.5,22){\circle*{2}}
\end{picture}
}
;\;
0001
=
\raisebox{-0.7cm}{
\begin{picture}(25,22)(0,3)
\put(5,5){\circle{2}}
\put(5,6){\line(0,1){8.5}}
\put(5,15){\circle{2}}
\put(20,15){\circle{2}}
\put(5,16){\line(1,1){6}}
\put(19.5,16){\line(-1,1){6}}
\put(12.5,22){\circle*{2}}
\end{picture}
}
\]
There are 3 equivalence classes of timelets on $\causet$ with respect to the monotonicity reduction \eqref{emonequiv}, corresponding to the following ordering of the elements of $\causet$:
\[
1234, 1324, 2134
\]
The normalized timelet associated with the first ordering looks like (the numbers next to the vertices show the values assigned):
\begin{equation}\label{eex1234}
\timelet_{1234}
\;=\;
\raisebox{-0.7cm}{
\begin{picture}(30,22)(0,3)
\put(5,5){\circle{2}}
\put(1,2){\mbox{\small{0}}}
\put(1,13){\mbox{\small{2}}}
\put(16,13){\mbox{\small{1}}}
\put(7,21){\mbox{\small{3}}}
\put(5,6){\line(0,1){8.5}}
\put(5,15){\circle{2}}
\put(20,15){\circle{2}}
\put(5,16){\line(1,1){6}}
\put(19.5,16){\line(-1,1){6}}
\put(12.5,22){\circle{2}}
\end{picture}
}
\end{equation}and decomposes as
\[
\timelet_{1234}
=
0111++0011+0001
\]
Another one decomposes according to eqref{edect} as follows:
\begin{equation}\label{eex1324}
\timelet_{1324}
\;=\;
\raisebox{-0.7cm}{
\begin{picture}(30,22)(0,3)
\put(5,5){\circle{2}}
\put(1,2){\mbox{\small{0}}}
\put(1,13){\mbox{\small{1}}}
\put(16,13){\mbox{\small{2}}}
\put(7,21){\mbox{\small{3}}}
\put(5,6){\line(0,1){8.5}}
\put(5,15){\circle{2}}
\put(20,15){\circle{2}}
\put(5,16){\line(1,1){6}}
\put(19.5,16){\line(-1,1){6}}
\put(12.5,22){\circle{2}}
\end{picture}
}
\; = \;
0111+0101+0001
\end{equation}
and the last one as
\begin{equation}\label{eex3124}
\timelet_{3124}
\;=\;
\raisebox{-0.7cm}{
\begin{picture}(30,22)(0,3)
\put(5,5){\circle{2}}
\put(1,2){\mbox{\small{1}}}
\put(1,13){\mbox{\small{2}}}
\put(16,13){\mbox{\small{0}}}
\put(7,21){\mbox{\small{3}}}
\put(5,6){\line(0,1){8.5}}
\put(5,15){\circle{2}}
\put(20,15){\circle{2}}
\put(5,16){\line(1,1){6}}
\put(19.5,16){\line(-1,1){6}}
\put(12.5,22){\circle{2}}
\end{picture}
}
\; = \;
1101+0101+0001
\end{equation}
Now we are in a position to draw the simplicial complex of timelets.
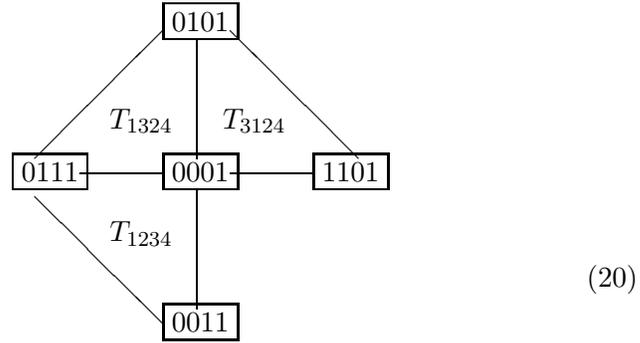
\unitlength1mm
\begin{equation}\label{eex1sim}
\raisebox{-.6cm}{
\begin{picture}(40,44)(0,0)
\put(20,0){\fbox{0011}}
\put(0,20){\fbox{0111}}
\put(20,20){\fbox{0001}}
\put(40,20){\fbox{1101}}
\put(20,40){\fbox{0101}}
\put(3,18){\line(1,-1){17}}
\put(24.5,3){\line(0,1){16}}
\put(9,21){\line(1,0){11}}
\put(29,21){\line(1,0){11}}
\put(3,23){\line(1,1){17}}
\put(24.5,23){\line(0,1){16}}
\put(46,23){\line(-1,1){17}}
\put(13,12){$\timelet_{1234}$}
\put(13,27){$\timelet_{1324}$}
\put(28,27){$\timelet_{3124}$}
\end{picture}
}
\end{equation}
There are three 2-simplices, each associated with appropriate equivalence class of timelets.
\paragraph{Example 2.} Consider the causet $\causet$ (the numbers next to the vertices are just labels) and write down the decomposition of generic timelets on $\causet$ in detail.
\unitlength1mm
\begin{equation}\label{eex2}
\causet
\;=\;
\raisebox{-.6cm}{
\begin{picture}(25,18)(0,3)
\put(5,5){\circle{2}}
\put(20,5){\circle{2}}
\put(5,15){\circle{2}}
\put(20,15){\circle{2}}
\put(1,2){\mbox{\small{1}}}
\put(16,2){\mbox{\small{2}}}
\put(1,13){\mbox{\small{3}}}
\put(16,13){\mbox{\small{4}}}
\put(5,6){\line(0,1){8.5}}
\put(20,6){\line(0,1){8.5}}
\end{picture}
}
\end{equation}
For this causet, the following nondegenerate coevents are preclusive (filled circle corresponds to value 1 on the appropriate element of $\causet$), here is the complete list of the elements of $\baryprec$:
\unitlength1mm
\[
0111
\;=\;
\raisebox{-.6cm}{
\begin{picture}(25,18)(0,3)
\put(5,5){\circle{2}}
\put(20,5){\circle*{2}}
\put(5,15){\circle*{2}}
\put(20,15){\circle*{2}}
\put(5,6){\line(0,1){8.5}}
\put(20,6){\line(0,1){8.5}}
\end{picture}
}
; \;
1011
\;=\;
\raisebox{-.6cm}{
\begin{picture}(25,18)(0,3)
\put(5,5){\circle*{2}}
\put(20,5){\circle{2}}
\put(5,15){\circle*{2}}
\put(20,15){\circle*{2}}
\put(5,6){\line(0,1){8.5}}
\put(20,6){\line(0,1){8.5}}
\end{picture}
}
\]
\[
0011
\;=\;
\raisebox{-.6cm}{
\begin{picture}(25,18)(0,3)
\put(5,5){\circle{2}}
\put(20,5){\circle{2}}
\put(5,15){\circle*{2}}
\put(20,15){\circle*{2}}
\put(5,6){\line(0,1){8.5}}
\put(20,6){\line(0,1){8.5}}
\end{picture}
}
; \;
0101
\;=\;
\raisebox{-.6cm}{
\begin{picture}(25,18)(0,3)
\put(5,5){\circle{2}}
\put(20,5){\circle*{2}}
\put(5,15){\circle{2}}
\put(20,15){\circle*{2}}
\put(5,6){\line(0,1){8.5}}
\put(20,6){\line(0,1){8.5}}
\end{picture}
}
; \;
1010
\;=\;
\raisebox{-.6cm}{
\begin{picture}(25,18)(0,3)
\put(5,5){\circle*{2}}
\put(20,5){\circle{2}}
\put(5,15){\circle*{2}}
\put(20,15){\circle{2}}
\put(5,6){\line(0,1){8.5}}
\put(20,6){\line(0,1){8.5}}
\end{picture}
}
\]
\[
0010
\;=\;
\raisebox{-.6cm}{
\begin{picture}(25,18)(0,3)
\put(5,5){\circle{2}}
\put(20,5){\circle{2}}
\put(5,15){\circle*{2}}
\put(20,15){\circle{2}}
\put(5,6){\line(0,1){8.5}}
\put(20,6){\line(0,1){8.5}}
\end{picture}
}
; \;
0001
\;=\;
\raisebox{-.6cm}{
\begin{picture}(25,18)(0,3)
\put(5,5){\circle{2}}
\put(20,5){\circle{2}}
\put(5,15){\circle{2}}
\put(20,15){\circle*{2}}
\put(5,6){\line(0,1){8.5}}
\put(20,6){\line(0,1){8.5}}
\end{picture}
}
\]
There are 6 equivalence classes of timelets on $\causet$ with respect to the monotonicity reduction \eqref{emonequiv}, corresponding to the following ordering of the elements of $\causet$:
\[
1234, 1243, 1324, 2134, 2143, 2413
\]
The normalized timelet associated with the first ordering has the following decomposition:
\[
\timelet_{1234}
\;=\;
\raisebox{-.6cm}{
\begin{picture}(21,18)(0,3)
\put(5,5){\circle{2}}
\put(20,5){\circle*{2}}
\put(5,15){\circle*{2}}
\put(20,15){\circle*{2}}
\put(5,6){\line(0,1){8.5}}
\put(20,6){\line(0,1){8.5}}
\end{picture}
}
\;+
\raisebox{-.6cm}{
\begin{picture}(21,18)(0,3)
\put(5,5){\circle{2}}
\put(20,5){\circle{2}}
\put(5,15){\circle*{2}}
\put(20,15){\circle*{2}}
\put(5,6){\line(0,1){8.5}}
\put(20,6){\line(0,1){8.5}}
\end{picture}
}
\;+
\raisebox{-.6cm}{
\begin{picture}(21,18)(0,3)
\put(5,5){\circle{2}}
\put(20,5){\circle{2}}
\put(5,15){\circle{2}}
\put(20,15){\circle*{2}}
\put(5,6){\line(0,1){8.5}}
\put(20,6){\line(0,1){8.5}}
\end{picture}
}
=0111+0011+0001
\]
Further timelets decompose as follows (the numbers next to the vertices show the values assigned):
\[
\timelet_{1243}
\;=\;
\raisebox{-.6cm}{
\begin{picture}(25,18)(0,3)
\put(5,5){\circle{2}}
\put(20,5){\circle{2}}
\put(5,15){\circle{2}}
\put(20,15){\circle{2}}
\put(5,6){\line(0,1){8.5}}
\put(20,6){\line(0,1){8.5}}
\put(1,2){\mbox{\small{0}}}
\put(16,2){\mbox{\small{1}}}
\put(1,13){\mbox{\small{3}}}
\put(16,13){\mbox{\small{2}}}
\end{picture}
}
\;=\;
0111+0011+0010
\]
\[
\timelet_{1324}
\;=\;
\raisebox{-.6cm}{
\begin{picture}(25,18)(0,3)
\put(5,5){\circle{2}}
\put(20,5){\circle{2}}
\put(5,15){\circle{2}}
\put(20,15){\circle{2}}
\put(5,6){\line(0,1){8.5}}
\put(20,6){\line(0,1){8.5}}
\put(1,2){\mbox{\small{0}}}
\put(16,2){\mbox{\small{2}}}
\put(1,13){\mbox{\small{1}}}
\put(16,13){\mbox{\small{3}}}
\end{picture}
}
\;=\;
0111+0101+0001
\]
\[
\timelet_{2134}
\;=\;
\raisebox{-.6cm}{
\begin{picture}(25,18)(0,3)
\put(5,5){\circle{2}}
\put(20,5){\circle{2}}
\put(5,15){\circle{2}}
\put(20,15){\circle{2}}
\put(5,6){\line(0,1){8.5}}
\put(20,6){\line(0,1){8.5}}
\put(1,2){\mbox{\small{1}}}
\put(16,2){\mbox{\small{0}}}
\put(1,13){\mbox{\small{2}}}
\put(16,13){\mbox{\small{3}}}
\end{picture}
}
\;=\;
1011+0011+0001
\]
\[
\timelet_{2143}
\;=\;
\raisebox{-.6cm}{
\begin{picture}(25,18)(0,3)
\put(5,5){\circle{2}}
\put(20,5){\circle{2}}
\put(5,15){\circle{2}}
\put(20,15){\circle{2}}
\put(5,6){\line(0,1){8.5}}
\put(20,6){\line(0,1){8.5}}
\put(1,2){\mbox{\small{1}}}
\put(16,2){\mbox{\small{0}}}
\put(1,13){\mbox{\small{3}}}
\put(16,13){\mbox{\small{2}}}
\end{picture}
}
\;=\;
1011+0011+0010
\]
\[
\timelet_{2413}
\;=\;
\raisebox{-.6cm}{
\begin{picture}(25,18)(0,3)
\put(5,5){\circle{2}}
\put(20,5){\circle{2}}
\put(5,15){\circle{2}}
\put(20,15){\circle{2}}
\put(5,6){\line(0,1){8.5}}
\put(20,6){\line(0,1){8.5}}
\put(1,2){\mbox{\small{2}}}
\put(16,2){\mbox{\small{0}}}
\put(1,13){\mbox{\small{3}}}
\put(16,13){\mbox{\small{1}}}
\end{picture}
}
\;=\;
1011+1010+0010
\]
The appropriate simplicial complex looks like follows:
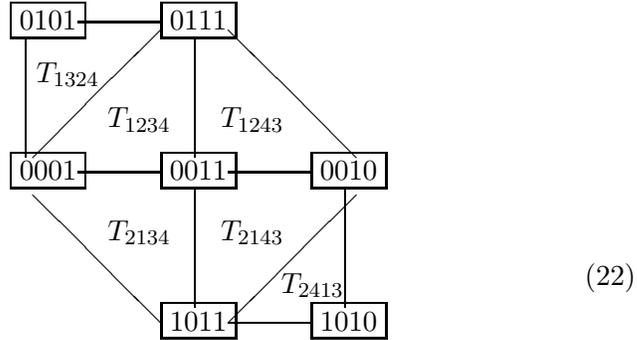
\unitlength1mm
\begin{equation}\label{eex2sim}
\raisebox{-.6cm}{
\begin{picture}(40,44)(0,0)
\put(20,0){\fbox{1011}}
\put(0,20){\fbox{0001}}
\put(20,20){\fbox{0011}}
\put(40,20){\fbox{0010}}
\put(20,40){\fbox{0111}}
\put(3,18){\line(1,-1){17}}
\put(24.5,3){\line(0,1){16}}
\put(46,18){\line(-1,-1){17}}
\put(9,21){\line(1,0){11}}
\put(29,21){\line(1,0){11}}
\put(3,23){\line(1,1){17}}
\put(24.5,23){\line(0,1){16}}
\put(46,23){\line(-1,1){17}}
\put(0,40){\fbox{0101}}
\put(2,39){\line(0,-1){16}}
\put(9,41){\line(1,0){11}}
\put(40,0){\fbox{1010}}
\put(29,1){\line(1,0){11}}
\put(44.5,3){\line(0,1){16}}
\put(3.5,33){$\timelet_{1324}$}
\put(13,12){$\timelet_{2134}$}
\put(13,27){$\timelet_{1234}$}
\put(28,27){$\timelet_{1243}$}
\put(28,12){$\timelet_{2143}$}
\put(36,5){$\timelet_{2413}$}
\end{picture}
}
\end{equation}

\section*{Afterword}

I introduced timelets as dual structures on causal sets. Intuitively, if you think of causets as spacetime models, they play the r\^ole of global time coordinates. Each equivalence class of timelets can be associated with a sequence of consecutive events, making it similar to homogeneous history, while the representation of all equivalence classes of timelets can be treated as a non-trivial geometric structure imposed on decoherent histories, where coevents are essentially employed \cite{Wallden2013}.


\begin{thebibliography}{99}

\bibitem{Burkill1984}
H.~Burkill,
Monotonic functions on partially ordered sets,
J. Comb. Theory, Ser. A,
{\bf 37}, 3, 248-256 (1984)

\bibitem{RideoutSorkin99}
D. P. Rideout, R. D. Sorkin, 
A Classical Sequential Growth Dynamics for Causal Sets
Phys.Rev.D61:024002,2000, 
arXiv:gr-qc/9904062

\bibitem{Sorkin2007}
Rafael D. Sorkin, 
An exercise in "anhomomorphic logic",
J. Phys. Conf. Ser.67:012018,2007,
arXiv:quant-ph/0703276

\bibitem{Wallden2013}
Petros Wallden,
The coevent formulation of quantum theory,
J. Phys. Conf. Ser. 442, 012044 (2013),
arXiv:1301.5704 [quant-ph]


\end{thebibliography}
\end{document}